\documentclass[aps,twocolumn,nopreprintnumbers,nolongbibliography,prb,10pt,showpacs]{revtex4-1}

\usepackage{bm}
\usepackage{url}
\usepackage{amsmath}
\usepackage{amssymb}
\usepackage{amsxtra}
\usepackage{amscd}
\usepackage{amsthm}
\usepackage{amsfonts}
\usepackage{eucal}
\usepackage{latexsym,amsthm}
\usepackage{pstricks}
\usepackage{color}
\usepackage{graphicx}

\newcommand{\nc}{\newcommand}
\nc{\on}{\operatorname}
\nc{\wt}{\widetilde}
\nc{\Wick}{{\mathbb :}}
\nc{\R}{{\mathbb R}}

\newcommand{\beq}{\begin{equation}}
\newcommand{\eeq}{\end{equation}}
\newcommand{\bmul}{\begin{multline}}
\newcommand{\emul}{{\end{multline}}}
\newcommand\beqa{\begin{eqnarray}}
\newcommand\eeqa{\end{eqnarray}}
\newcommand\bea{\begin{array}}
\newcommand\eea{\end{array}}
\newcommand\ba{\begin{array}}
\newcommand\ea{\end{array}}
\newcommand{\nn}{\nonumber}

\newcommand{\neqa}{\nonumber\end{eqnarray}}

\newcommand{\eq}[1]{Eq.(\ref{#1})}

\renewcommand{\d}{\partial}

\nc{\CH}{{\mathcal H}}
\nc{\Db}{{\bar D}}
\nc\comment[1]{}

\nc{\CM}{{\mathcal M}}
\nc{\CN}{{\mathcal N}}

\newcommand{\re}{\relax{\rm I\kern-.18em R}}

\nc{\meV}{{\mathrm{\,meV}}}
\nc{\cG}{{\mathcal G}}

\renewcommand{\bar}{\overline} 

\nc{\al}{{\alpha}}

\def\eps{{\epsilon}}
\def\sign{{\rm \, sign }}

\renewcommand{\)}{\right)}
\renewcommand{\(}{\left(}

\begin{document}
\comment{$}
\title{Charging of graphene by magnetic field and mechanical effect of magnetic oscillations}
\author{Sergey Slizovskiy}
\email{S.Slizovskiy@lboro.ac.uk}
\affiliation{Department of Physics,  Loughborough University,\\
Loughborough LE11 3TU, UK}
\affiliation{Petersburg Nuclear Physics Institute, Russia}

\begin{abstract}
We discuss the fact that quantum capacitance of graphene-based devices leads to variation of graphene charge density under
changes of external magnetic field. The charge is conserved, but redistributes to substrate or other graphene sheet. 
We derive exact analytic expression for charge redistribution in the case of ideal graphene in strong magnetic field. 
When we account for impurities and temperature, the effect decreases and the formulae reduce to standard quantum capacitance
expressions. The importance of quantum capacitance for
potential Casimir force experiments is emphasized and the corresponding corrections are worked out. 
\end{abstract}

\maketitle

{\it Introduction.}
Graphene is a novel two-dimensional material having unique mechanical and electronic properties.  The uniqueness of any two-dimensional
material is that it's electronic properties can be easily tuned by doping or gating. On top of that, 
graphene is the strongest known
Quantum Hall material due to a sharp conical tip of it's linear dispersion near the Dirac point. 
Magnetic oscillations can be noticed
even at room temperature for magnetic field
$B\gtrsim 5$ T \cite{Review09,QCapacitance10}.
 
The electronic properties of materials show up in the Casimir effect. There was a flurry of recent theoretical activity devoted to
computation of Casimir effect for graphene \cite{Fialkovsky09,Fialkovsky11,Manybody11,TwoModels12,Macdonald12}, with several controversies
still unresolved, and so it is important to
perform experiments to verify these computations. 
It is hard to make a mechanical measurement of Casimir force in graphene due to it's two-dimensional nature and since it is almost
always electrically charged.  The electrostatic force is much stronger than the Casimir force, so,
one needs to subtract the electrostatic contribution to single out the fluctuation-induced Casimir force.
The first experiments have started to appear only recently \cite{Measurement13}. Since graphene exhibits a strong Quantum Hall Effect
(QHE) it would be of interest to repeat the experiments \cite{Measurement13} with strong transverse magnetic field.

A method for subtracting the (clearly dominant) electrostatic force \cite{Subtraction09} was used in \cite{Measurement13}: the
electrostatic force depends on the gate voltage $V$ as $F \sim (V-V_0)^2$, where $V_0$ is a residual graphene voltage due to charged
impurities and chemical potential difference with substrate. 
The formula
has allowed the authors of \cite{Measurement13} to find the gate voltage $V_0$ where electrostatic force is fully compensated. 
The above formula does not include the quantum capacitance contribution
and, which is included by adding the ``quantum capacitor'' in a series connection: 
$ \sigma = \int_{V_0}^V c_{total}(V') dV' = \int_{V_0}^V (c^{-1} + c_q^{-1}(V'))^{-1} dV'$ ($\sigma$ is a charge density, 
$c=\eps/d$ is
geometric capacitance per unit area and $c_q(V') = e^2 D(\mu_g(V'),T)$, where $D = dn/dE$ is a density of
states \cite{QCapacitance88,QCapacitance94,QCapacitance10} is quantum capacitance). 
Then the electric pressure is $F/Area = \sigma^2/(2 \eps)$, so we 
get
\beqa \label{ForceVariation}
 & F/Area = \frac{1}{2 \eps} \left( \int_{V_0}^V (c^{-1} + c_q^{-1}(V'))^{-1} dV'\right )^2 \\ 
 & \nn \approx  \frac{1}{2 \eps} \left( (V_0 - V)^2 c^2 - 2 (V_0 - V) c^4 \int_{V_0}^V \frac{dV'}{e^2 D(\mu_g(V'),T)}\right)      
\eeqa
where $D(\mu,T) = \int f'(E-\mu,T) D(E) dE$,  $f(\mu,T)$ is a Fermi-Dirac distribution,  $D(\mu)$ is a density of states for
graphene. The chemical potential $\mu_g(V)$ for graphene depends on the applied voltage and the chemical doping may give
a constant shift: $\mu_g(V) = - e\, V + const$.  
 For ideal graphene $D(E)= \frac{2 E}{\pi v_F^2}$
and so $D^{-1}(E,T)$ gives a singular contribution near the Dirac cone ($E=0$) at small temperatures. Due to charge puddles in 
realistic graphene on substrate, the inverse density of states becomes smooth in the vicinity of the Dirac 
point \cite{QCapacitance10}, hence it gives
a weakly $V$-dependent quantum capacitance of order $10^{-2} {\rm F/m^2}$, thus the simple $(V-V_0)^2$ fit should work
well for small intervals of $V-V_0$.
 
The story gets more interesting with magnetic field. The Casimir force for this case was estimated in Ref.\onlinecite{Macdonald12}, where
pronounced dependence on the magnetic field and the chemical potential was shown. Thus it makes sense to scan 
a wider range of chemical potentials in the experiment.  The magnetic field does
also influence the electrostatic force, since the charge of ideal graphene is a step-like function of chemical potential with size of the
step depending on the magnetic field value. Thus, even if we consider a suspended graphene with only chemical doping, its charge will
oscillate when changing magnetic field. 
The discussion of electrostatic
contribution in magnetic field and quantum capacitance effect is the aim of this note.
 
    
Below we consider three examples:
\begin{itemize}  
\item Graphene  suspended over the wide
trench etched in a metallic substrate, or, alternatively, it can be suspended by leaning on crests. 
\item Two sheets of graphene forming a capacitor with fixed voltage
applied (such geometry was discussed in Ref.\onlinecite{Macdonald12} and argued to have a possibility of repulsive Casimir force)
\item Graphene laying on the insulator-coated semiconductor with given gate voltage $V_{gate}$ and with grounded parallel metallic plate
(or sphere)
hanging at the distance $d$ over graphene (actually, it is attached to vibrating cantilever of atomic-force microscope). 
Such geometry was
used in the recent experiment \cite{Measurement13}.
\end{itemize}
Below we derive explicit analytic formulas for the case of ideal graphene at zero temperature and then discuss a more realistic
situation with the approach similar to Ref.\onlinecite{QCapacitance10,NonlinearMagnetization}.  

 With magnetic field the energy levels of conduction band of graphene are  $E_k=\sign(k) \sqrt{\alpha |k B|}$ where $\alpha = 2 \hbar e
v_F^2$ 
and $k$ is an integer,  with degeneracy $g_s g_v C |B|$ per unit area, where $C=\frac{e}{2 \pi \hbar}$ and  
$g_v=2$, $g_s=2$ accounts for
spin and valley degeneracy.   

 When the levels are quantized, only the levels below the chemical potential $\mu_{g}$ would be filled. For undoped graphene one would
have half-filled zero LL, this serves as a reference point for summation of formally infinite spectrum of ``Dirac sea''.
 For generic $\mu$ the charge density is quantized and given by 
\beq \label{n(B)}
 n(B) = 4 C |B| \(\left[\frac{\mu_{g}^2 \sign(\mu_g)}{\alpha |B|}\right]+1/2\)
\eeq
where $[]$ denotes the integer part (Floor). 

Consider {\it the case of graphene  suspended over the etched
trench of depth $d$ in a metallic substrate}. In this geometry graphene is connected
to a conductor. Another example to which the same computation applies is  a piece of pyrographite from which
a large graphene flake has exfoliated. 

Consider graphene having the chemical potential $\mu_{g0}$ for mobile carriers with density $n_0$ at zero magnetic field.
These are related as
\beq
 n_0=\frac{1}{\pi} \sign(\mu_{g0}) \(\frac{\mu_c}{\hbar v_F}\)^2 = 4 \sign(\mu_{g0}) \frac{C}{\alpha} \mu_{g0}^2
\eeq

\begin{figure}[t]  
\begin{center}
\includegraphics[scale=0.8]{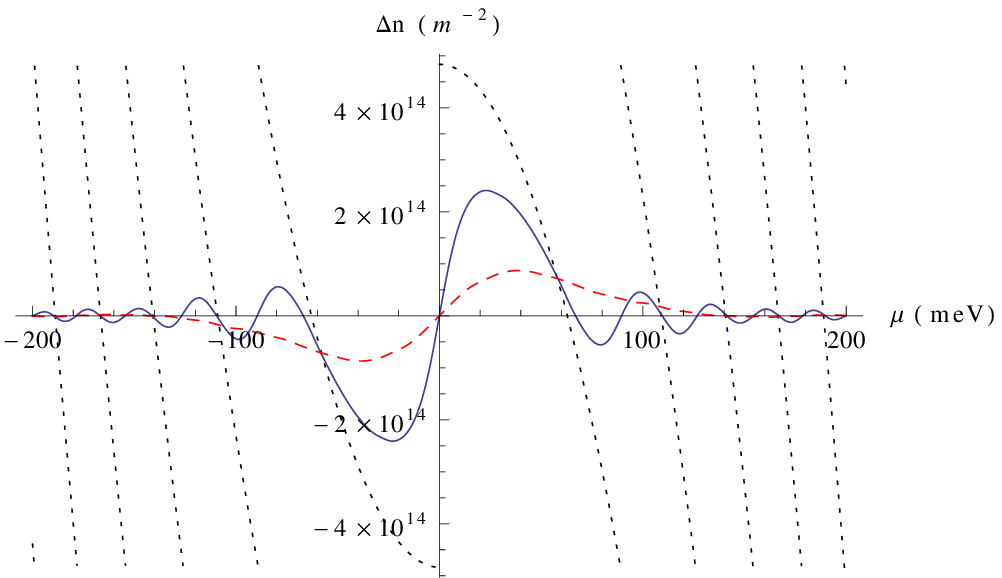}
\includegraphics[scale=0.8]{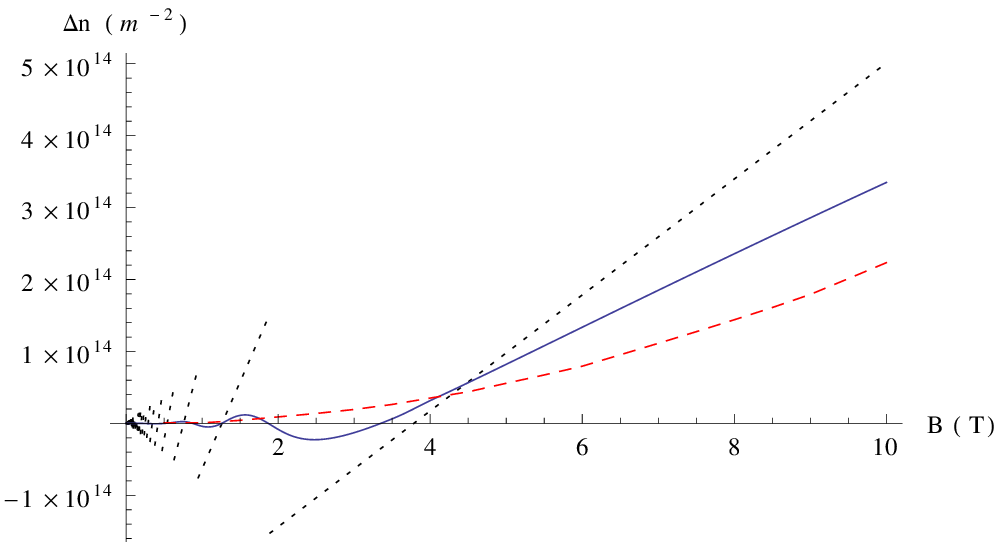}
\end{center}
\caption{\label{fig:FreeCharging} 
Change of electron density ($\rm{m^{-2}}$) when turning on the magnetic field (B = 6 T) as a function of initial 
chemical potential (first plot) and as a function of magnetic field
(for
$\mu_{graphene} = 50 \meV$), 
capacitance denominator is set here to 6. Dotted curves are for ideal graphene, solid curve accounts for 
Lorentzian broadening $5 \sqrt{B ({\mathrm T})} \meV$, dashed curve corresponds to broadening due to temperature $T=300 K$.}
\end{figure}

When the magnetic field is switched on, the electronic structure of graphene changes much stronger than the one for the other
materials involved, so, we consider the effect of magnetic field only on graphene and thus the chemical potential of 
the conductor in the
bottom of the trench is fixed (here we neglect the electric penetration depth for the conductor). 
Since the magnetic field may induce
changes of the carrier number of graphene, $n=n_0 + \delta n$, this creates an extra electric field $\delta E = - e \delta n/\epsilon$
which shifts the chemical potential of graphene by $-e d \delta E$, so  we
solve 
\beq
\mu_{g}-\mu_{g0} =  e^2 \delta n/c
\eeq
where $\delta n$ depends on $B$ and
$
c =  \eps/d 
$
is a capacitance per unit area: $d$ is the distance between the plates of the capacitor and $\eps$ is a dielectric permittivity 
($\eps = \eps_0$ for the vacuum).
  Using \eq{n(B)} we get the equation:
\beqa
&4 C e^4 \sign(\mu_{g0})\delta n^2 + ( \al c^2 + 8 C c e^2 |\mu_{g0}|) \delta n + \nn \\ 
&+ 4 C c^2 |B| \al \(\left\{\frac{\mu_{g0}^2
\sign(\mu_{g0})}{\alpha |B|}
\right\}-\frac12 \)=0 \ ,
\eeqa which has a simple solution in the limit  $\delta n \ll n_0 $ :
\beq \label{Charge Oscillations}
\delta n = \frac{4 C |B|}{1+ 8 C |\mu_{g0}| e^2/(\alpha c)} \(\frac12  - \left\{\frac{\mu_{g0}^2 \sign(\mu_{g0})}{\alpha
|B|} 
\right\}\) \eeq
where $\{x\} = x - [x]$ gives the fractional part. The exact solution is also straightforward.
This result shows how the charge of graphene oscillates when the magnetic field is changed, see the dotted curve in
Fig.\ref{fig:FreeCharging}.  The corresponding force oscillation follows from $F/Area = e^2 n^2/(2 \eps)$. 

It is clear that temperature and disorder would reduce the effect we discuss. 
For the case of very clean suspended graphene we expect the disorder to be weak and choose a simplified model of equal-shape broadening of
all the Landau levels. It is clear that the actual result would depend 
mostly of the shape of the level that is nearest to $\mu_{g0}$, so, it's the width of that level that we should take as our
broadening.  The broadening is computationally equivalent to smearing of the chemical potential,
see Fig. \ref{fig:FreeCharging}.

The effect we discuss is another manifestation of integer Quantum Hall Effect.
Qualitatively, if the last
filled Landau level (LL) is less than half-filled, then the chemical potential is higher than the one without magnetic field, so, graphene
wants to get rid of carriers and gets positively charged;  the opposite happens for more-than-half filled level. This also shows that
the upper bound for magnetic charging of graphene is half the population of one LL: $\frac12 4 C |B|$, this bound is never achieved
due to non-infinite geometric capacitance and level broadening. 
For example, with $d=10$nm and $\mu_{g0} = 20 \meV$ we get in the denominator:
$1+ 8 C |\mu_{c}| e^2/(\alpha c) = 6.3 $, which is a typical quantum to geometric capacitance ratio for graphene experiments on
thin insulator layers.

The magnetic oscillations of charge 
have a mechanical effect, creating the attraction between the plates of charged capacitor.  
We see that a typical variation of electron density could be of order $10^{15} \rm{m^{-2}}$, which translates into the electric pressure
\beq
P = n^2 e^2/(2 \eps_0)  \sim 2000\,{\rm Pa}
\eeq
This pressure is of the same order of magnitude as the Casimir pressure at distance $10$ nm between the plates (
estimated to be roughly $10^{-3} \hbar c \pi^2/(240 d^4)$.  Note that the 
electrostatic force from the magnetic charging effect falls off as $1/d^3$ due to linearly decreasing capacity, while
the Casimir force falls off as $1/d^4$ for small temperatures, so these effects are comparable.

 Let us see if it is feasible to measure the effect. Consider a trench of width  $l=100$ nm.
Then the membrane has a parabolic form with tension $T = 1/2 \sqrt{k l P}$ and the central deflection is 
$h_0 = l^2 P/T/8 = l^{3/2} P^{1/2} k^{-1/2}$ where $k  \approx 300 \rm{N/m}$ is a 2D Young modulus
(note that it is possible that for small deformations the Hookes law is invalid for graphene due to
microscopic out of plane buckled form of graphene \cite{Kusmartsev12, Kusmartsev12-1}, thus, for small deformations the effective Young
modulus could be lower). For
$l=100$nm  trench we get $h_0 \approx 10^{-11} m$, which can be measured in STM or in Bragg diffraction experiments. For wider
trenches the deflection grows as $l^{3/2}$. 

Having the possibility to measure the electric attraction, one can apply voltage to graphene and tune it to minimize the attraction,
analogously to Ref.\onlinecite{Measurement13}. 

Now consider a geometry of {\it two parallel sheets of graphene}, that are electrically connected. This may be 
imagined as a drum made of two grapehene sheets. 
Let these sheets be doped to $\mu_{10}$ and $\mu_{20}$ without magnetic field and have carrier densities $n_{01}, n_{02}$.
An interest in such type of geometry stems from the prediction 
of possible repulsive Casimir force in magnetic field when $\mu_{1}$ and $\mu_{2}$ are of opposite signs \cite{Macdonald12}.  
In magnetic field we assume carrier density redistribution $n_{1,2}(B) = n_{0\,1,2} \pm \delta n(B)$ and solve:
\beq
\mu_1-\mu_{10} = \mu_2-\mu_{20} + e^2 \delta n/c
\eeq   
together with \eq{n(B)} for both graphene sheets. 
The solution in the approximation $\delta n \ll n$ is
\beqa
&\delta n  = 4 C |B| \times \\ \nn & \times\frac{|\mu_{10}| \(\left\{ \frac{\mu_{20}^2 \sign(\mu_{20})}{\al |B|} \right\} -
\frac12 \)
-
|\mu_{20}| \(\left\{ \frac{\mu_{10}^2 \sign(\mu_{10})}{\al |B|} \right\} - \frac12 \) }{|\mu_{10}|+|\mu_{20}| + 8 C |\mu_{10}
\mu_{20}| e^2/(\al c)}
\eeqa
where $c =  \eps/d$.  Note that the effect of magnetic oscillations cancels out if the two graphene sheets are at equal chemical
potentials (and are of equal quality).

Now we turn to a much more flexible experimental setup used in \cite{Measurement13} and discuss {\it gated graphene laying on the
insulator-coated semiconductor with a grounded parallel metal plate (or sphere) hanging over
it}, the upper plate is an atomic force microscope used in frequency-shift regime \cite{Method12,Method121,Measurement13}. The
presence of substrate and a larger distance to the 
metal plate (of order 300 nm) makes the quantum capacitance effects much weaker, but these are still important to
improve precision. Remarkably,  {\it  this experimental setup allows for an excellent direct mechanical measurement of
magnetic oscillations together with QHE.} 

Now we assume only weak magnetic oscillations, so the density of states is a smooth function and it is convenient to reformulate
the solution in terms of continuum density of states: we have a series connection of two capacitors: the standard geometric one with 
$c_{geom}=\eps/d$ (per unit area) and a quantum one with $c_q = e^2 D(\mu_g)$, where $D = dn/dE$ is a density of
states \cite{QCapacitance88,QCapacitance94,QCapacitance10}. So, the total capacitance is $c=(c_{geom}^{-1} + c_q^{-1})^{-1}$.  We see
that
the relative effect of
quantum capacitance decreases as $1/d$ due to decreasing of $c_{geom}$, so, for fixed voltage its contribution to force decreases as
$1/d^3$, which is small, but may still compete with the Casimir force that behaves as $1/d^4$ at low temperatures \cite{Macdonald12}.

To study magnetic oscillations and Casimir effect at strong magnetic field one needs to extend the experiment of
Ref.\onlinecite{Measurement13} by extra bottom-gating, so that a wide range of Landau level filling factors could be scanned. 

\begin{figure}[t]  
\begin{center}
\includegraphics[scale=0.78]{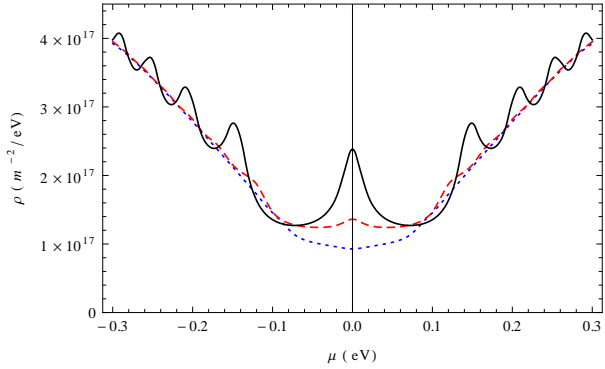}
\end{center}
\vspace{-0.6 cm}
\caption{\label{fig:DOS} 
Density of states with Lorentz broadening 15 meV and carrier concentration dispersion $4 \cdot 10^{15} \rm{m^{-2}}$ as a function of
chemical potential for $B=$ 5 T (dotted), 10 T(dashed), 16 T(solid).}
\end{figure}

For the electrostatic force acting on the unit area of graphene we may use \eq{ForceVariation} and  follow the model of Ref.
\onlinecite{QCapacitance10} to get the density of states. The model
consists of Lorentz and temperature level broadening superimposed on the Gaussian carrier number broadening due to charge puddles, see
Fig. \ref{fig:DOS}. 

With the improved fit the value of residual potential difference $V_0$ can be mechanically measured with fabulous precision.
$V_0$ is the potential difference between graphene and a metal plate when there is no electric field between them, so, it equals
to graphene chemical potential:
\beq
  V_0 = \mu_{graphene} + const
\eeq 
The electron doping of graphene is a linear function of bottom gate voltage (one can also easily write the 
quantum capacitance correction, but it is small for relatively thick insulator layer): $n \sim V_{gate}$.
So, the experiment allows for precise measurement of both $n$ and $\mu$. 
Knowing this for the particular sample is also helpful for theoretical refinement of Casimir force
computations. Importantly, the known $n(\mu,B)$ can be plugged back into \eq{ForceVariation} ($D = e^2 \frac{\d n}{\d \mu_{graphene}}$)
to improve the fit and hence the precision. 
To get a more pronounced Quantum Hall physics, the above experiment may be repeated with gated graphene suspended over thin layer of
insulator. Then one may hope to get a strong evidence for interaction effects. 

{\it To conclude}, we have elaborated on the two possible experimental schemes to measure the Casimir effect for graphene with magnetic
field. 
As a by-product, we note that the newly-developed mechanical method \cite{Method12,Method121,Measurement13} may lead to the precise
measurement of density of states if sample is additionally gated.  

{\it Acknowledgements:}
  I am grateful to Pablo Rodriguez-Lopez,  Ignat Fialkovsky, Galina Klimchitskaya and  Feo Kusmartsev for useful discussions.
This work has been supported by EPSRC through the grant EP/l02669X/1.

\bibliography{MyReferences}

\end{document}